\documentclass[twocolumn,showpacs,aps,prl,superscriptaddress]{revtex4}
\usepackage{graphicx}
\usepackage{dcolumn}
\usepackage{epsfig}
\usepackage{amsmath}
\RequirePackage{xspace}

\newcommand{\BaBarYear}    {06}
\newcommand{\BaBarNumber}  {012}
\newcommand{\SLACPubNumber} {11780}
\def\epem  {\ensuremath{e^+e^-}\xspace}
\newcommand\dbline{\noalign{\vskip 0.10truecm\hrule}\noalign{\vskip 2pt}\noalign{\hrule\vskip 0.10truecm}}
\providecommand{\tbline}{\noalign{\vskip 0.05truecm\hrule\vskip0.05truecm}}

\newcommand{\bma}[1]{\boldmath{$#1$}}
\providecommand{\bfemsix}{${\cal B} (10^{-6})$}

\newcommand{\thetaT}{\ensuremath{\theta_{\rm T}}}
\newcommand{\costhr}{\ensuremath{\cos\thetaT}}

\newcommand{\mres}{\ensuremath{m_{\rm res}}}
\def\jpsi     {\ensuremath{{J\mskip -3mu/\mskip -2mu\psi\mskip 2mu}}\xspace}
\def\ra                 {\ensuremath{\rightarrow}\xspace}

\newcommand{\etapr}{\ensuremath{\eta^{\prime}}\xspace}
\newcommand{\etapepp}{\ensuremath{\etapr_{\eta\pi\pi}}\xspace}
\newcommand{\etaprg}{\ensuremath{\etapr_{\rho\gamma}}\xspace}

\newcommand{\fetaprkgamma}{\ensuremath{\etapr K \gamma}}
\newcommand{\fetaprkpgamma}{\ensuremath{\etapr K^+ \gamma}}

\newcommand{\fetakgamma}{\ensuremath{\eta K \gamma}}
\newcommand{\fetakpgamma}{\ensuremath{\eta K^+ \gamma}}

\newcommand{\fetaggkpgamma}{\ensuremath{\eta_{\gamma \gamma} K^+ \gamma}}

\newcommand{\fetatrepikpgamma}{\ensuremath{\eta_{3\pi} K^+ \gamma}}

\newcommand{\fetaprRgkpgamma}{\ensuremath{\etapr_{\rho \gamma} K^+ \gamma}}
\newcommand{\fetaprEppkpgamma}{\ensuremath{\etapr_{\eta\pi\pi} K^+ \gamma}}

\newcommand{\fetaprkzgamma}{\ensuremath{\etapr K^{0} \gamma}}

\newcommand{\fetakzgamma}{\ensuremath{\eta K^{0} \gamma}}

\newcommand{\fetaggkzgamma}{\ensuremath{\eta_{\gamma \gamma} K^{0} \gamma}}

\newcommand{\fetatrepikzgamma}{\ensuremath{\eta_{3\pi} K^{0} \gamma}}

\newcommand{\fetaprRgkzgamma}{\ensuremath{\etapr_{\rho\gamma} K^{0} \gamma}}

\newcommand{\fetaprEppkzgamma}{\ensuremath{\etapr_{\eta\pi\pi} K^{0} \gamma}}

\providecommand{\BetapKGamma}{\mbox{$B \rightarrow \eta^{\prime} K \gamma $}}
\providecommand{\BetapKpGamma}{\mbox{$B^+ \rightarrow \eta^{\prime} K^+ \gamma $}}
\providecommand{\BetapKzGamma}{\mbox{$B^0 \rightarrow \eta^{\prime} K^0 \gamma $}}

\providecommand{\BetaKGamma}{\mbox{$B \rightarrow \eta K \gamma $}}
\providecommand{\BetaKpGamma}{\mbox{$B^+ \rightarrow \eta K^+ \gamma $}}

\newcommand{\etaKzg}{\mbox{$B^0\ra  \eta K^0  \gamma$}}
\newcommand{\etaKpg}{\mbox{$B^+\ra  \eta K^+  \gamma$}}

\newcommand{\etapKzg}{\mbox{$B^0\ra  \eta^{\prime} K^0  \gamma$}}
\newcommand{\etapKpg}{\mbox{$B^+\ra  \eta^{\prime} K^+  \gamma$}}

\newcommand{\psfile}[3][]{ 
  \begin{center}
    \setlength{\epsfxsize}{#3\linewidth}\leavevmode
    \def\noOpt{}\def\testit{#1}\ifx\testit\noOpt%
      \epsfbox{#2}%
    \else%
      \epsfbox[#1]{#2}%
    \fi
  \end{center}
}
\def\gaga  {\ensuremath{\gamma\gamma}\xspace} 
\def\pip   {\ensuremath{\pi^+}\xspace}
\def\pim   {\ensuremath{\pi^-}\xspace}
\def\piz   {\ensuremath{\pi^0}\xspace}

\newcommand{\etagg}{\ensuremath{\eta_{\gaga}}}
\newcommand{\etappp}{\mbox{$\eta_{3\pi}$}}
\newcommand{\rhoz}{\mbox{$\rho^0$}}

\def\KS    {\ensuremath{K^0_{\scriptscriptstyle S}}}

\def\Bbar    {\overline{B}{}}

\def\Bz      {\ensuremath{B^0}}

\newcommand{\DE}{\ensuremath{\Delta E}}
\def\mes{\mbox{$m_{\rm ES}$}}
\newcommand{\calB}{\mbox{${\cal B}$}}
\newcommand{\acp}{\ensuremath{{\cal A}_{ch}}}

\def\babar{{\em B}{\footnotesize\em A}{\em B}{\footnotesize\em AR}}
\def\pep2{PEP-II}

\newcommand{\UfourS}{\mbox{$\Upsilon(4S)$}}

\newcommand\etal{{\it et al.}}
\newcommand{\dedx}{\ensuremath{\mathrm{d}\hspace{-0.1em}E/\mathrm{d}x}}
\newcommand{\gevcc}{\mbox{$\textrm{GeV}/c^2$}} 
\newcommand{\mevcc}{\mbox{$\textrm{MeV}/c^2$}} 
\newcommand{\gevc}{\mbox{$\textrm{GeV}/c$}} 
\newcommand{\gev}{\mbox{$\textrm{GeV}$}} 
\newcommand{\mev}{\mbox{$\textrm{MeV}$}} 
 
\newcommand{\epjBase}        {Eur.\ Phys.\ Jour.\xspace}
\newcommand{\jprlBase}  [1]     {Phys.\ Rev.\ Lett. \xspace}
\newcommand{\jprl}      [1]    {\jprlBase\  ~{\bf #1}}
\newcommand{\jprBase}        {Phys.\ Rev.\ }
\newcommand{\jprd}      [1]  {\jprBase\ D~{\bf #1}}
\newcommand{\plBase}   [1]         {Phys.\ Lett. \xspace}
\newcommand{\plb}      [1]    {\plBase\  B~{\bf #1}}
\newcommand{\nimBaseA}       {Nucl.\ Instr.\ Meth.\ }
\newcommand{\nima}      [1]  {\nimBaseA~A~{\bf #1}}
\newcommand{\zpBase}         {Z.\ Phys.}
\newcommand{\zpc}       [1]  {\zpBase\ C~{\bf #1}}
\newcommand{\npBase}         {Nucl.\ Phys.\ }
\newcommand{\npb}       [1]  {\npBase\ B~{\bf #1}}

\newcommand{\jrmp}      [1]  {{Rev.\ Mod.\ Phys.\ {\bf #1}}}

\def\BB      {\ensuremath{B\Bbar}\xspace} 
\def\CP                {\ensuremath{C\!P}\xspace}
\newcommand{\half}{\mbox{${1\over2}$}}
\newcommand{\pvec}{{\bf p}}
\def\qqbar{\mbox{$q\bar q\ $}}
\newcommand{\xf}{\mbox{${\cal F}$}}
\newcommand{\signf}{$\cal S$ ($\sigma$)}

\newcommand{\eff}{$\epsilon$ (\%)}

  \newcommand{\BretaKzg}{\mbox{$\calB(\etaKzg)$}}
  \newcommand{\RetaKzg}{\ensuremath{11.3^{+2.8}_{-2.6}\pm 0.6}}       
  \newcommand{\setaKzg}{\ensuremath{5.3}}                    

  \newcommand{\BretaKpg}{\mbox{$\calB(\etaKpg)$}}
  \newcommand{\RetaKpg}{\ensuremath{10.0 \pm 1.3 \pm 0.5}}       
  \newcommand{\setaKpg}{\ensuremath{10.0}}                    
  \newcommand{\aetaKpg}{\ensuremath{-0.09 \pm 0.12 \pm 0.01}} 

  \newcommand{\BretapKzg}{\mbox{$\calB(\etapKzg)$}}
  \newcommand{\RetapKzg}{\ensuremath{1.1^{+2.8}_{-2.0}\pm 0.1}}       
  \newcommand{\uletapKzg}{\ensuremath{6.6}}                   
  \newcommand{\setapKzg}{\ensuremath{0.6}}                    

  \newcommand{\BretapKpg}{\mbox{$\calB(\etapKpg)$}}
  \newcommand{\RetapKpg}{\ensuremath{1.9^{+1.5}_{-1.2}\pm 0.1}}       
  \newcommand{\uletapKpg}{\ensuremath{4.2}}                   
  \newcommand{\setapKpg}{\ensuremath{1.7}}                    

\def\figurebox#1#2#3{%
    \def\arg{#3}%
    \ifx\arg\empty
    {\hfill\vbox{\hsize#2\hrule\hbox to #2{\vrule\hfill\vbox to #1{\hsize#2\vfill}\vrule}\hrule}\hfill}%
    \else
    {\hfill\epsfbox{#3}\hfill}%
    \fi}

\begin{document}
 

\preprint{\babar-PUB-\BaBarYear/\BaBarNumber} 
\preprint{SLAC-PUB-\SLACPubNumber} 

\begin{flushleft}
\babar-PUB-\BaBarYear/\BaBarNumber \\
 SLAC-PUB-\SLACPubNumber\\
\end{flushleft}

\title{\large  \bf\boldmath  Measurement of Branching Fractions in
  Radiative  $B$ Decays to \fetakgamma\  and Search for $B$ Decays to
  \fetaprkgamma\ }

%
\author{B.~Aubert}
\author{R.~Barate}
\author{M.~Bona}
\author{D.~Boutigny}
\author{F.~Couderc}
\author{Y.~Karyotakis}
\author{J.~P.~Lees}
\author{V.~Poireau}
\author{V.~Tisserand}
\author{A.~Zghiche}
\affiliation{Laboratoire de Physique des Particules, F-74941 Annecy-le-Vieux, France }
% ...
\collaboration{The \babar\ Collaboration}
\noaffiliation

\date{\today}

\begin{abstract}
We present  measurements of the \BetaKGamma\ branching fractions and 
upper limits  for the  \BetapKGamma\  branching fractions. For 
\BetaKpGamma\  we also measure  the time-integrated charge asymmetry.
 The data sample, collected with the \babar\ detector
at the Stanford Linear Accelerator Center, represents $232 \times 10^6$
produced  \BB\ pairs.
The results for branching fractions  
and upper limits at 90\% C.L. in units of $10^{-6}$ are:
$\BretaKzg = \RetaKzg$,
$\BretaKpg = \RetaKpg$,
$\BretapKzg <\uletapKzg$,
$\BretapKpg <\uletapKpg$.
The  charge asymmetry in the decay \etaKpg\ is $\acp = \aetaKpg$. The
first errors    are statistical and the second systematic.  
\end{abstract}


\maketitle

Radiative $B$ meson decays have long been recognized  as
a sensitive probe to test the Standard Model (SM) and to look 
for new physics (NP)~\cite{Hou} . 
In the SM, flavor-changing neutral current processes such as 
$b \rightarrow s \gamma$ proceed via radiative loop (penguin) diagrams.
The loop dia-grams may also contain  
new  heavy particles, and therefore are sensitive to NP.
Measurements of the branching fractions of a few of the
exclusive decay modes exist: $K^*(892) \gamma$ \cite{Cleo,Kstar}, $K_1(1270) \gamma$ \cite{yang},
$K^*_2(1430) \gamma$~\cite{Cleo,Kstar2}, $K \pi \pi \gamma$~\cite{Kstar2}, $\phi K \gamma$~\cite{KPhi} and      
$K^+ \eta \gamma$~\cite{KEta}.   The measured 
branching fraction of inclusive $b\rightarrow s \gamma$ 
and exclusive radiative $B$ decays are in good agreement with SM
predictions~\cite{PDG2004,ThPred}.  
Direct~\cite{Grueb} and mixing-induced~\cite{Gronau} \CP\ asymmetries
in exclusive radiative $B$ decays are expected to be very small in the SM. 
Measurement of direct \CP\ asymmetries in
exclusive radiative decays can provide a clear sign of NP~\cite{Atwood}. 

We present analyses of the exclusive decay modes
 \etaKpg\ and \etaKzg\ ~\cite{ChargeCon} , which have previously been
 measured  by the Belle Collaboration~\cite{KEta}, and 
\BetapKpGamma\  and \BetapKzGamma\ which are studied for the first time.
The results presented here are based on data collected
with the \babar\ detector~\cite{BABARNIM}
at the PEP-II asymmetric-energy $e^+e^-$ collider~\cite{pep}
located at the Stanford Linear Accelerator Center.
The analyses use an integrated
luminosity of 211~fb$^{-1}$, corresponding to 
$232$ million \BB\ pairs, recorded at the $\Upsilon (4S)$ 
resonance (at a center-of-mass energy of $\sqrt{s}=10.58\ \gev$).

Charged particles from \epem\ interactions are detected, and their
momenta measured, by a combination of a vertex tracker (SVT) consisting
of five layers of double-sided silicon microstrip detectors, and a
40-layer central drift chamber (DCH), both operating in the 1.5 T magnetic
field of a superconducting solenoid. We identify photons and electrons 
using a CsI(Tl) electromagnetic calorimeter (EMC).
Further charged-particle identification is provided by the average energy
loss (\dedx ) in the tracking devices and by an internally reflecting
ring-imaging Cherenkov detector (DIRC) covering the central region.
A $K/\pi$ separation of better than four standard deviations 
is achieved for momenta below 3~\gevc, decreasing to 2.5$\sigma$ at the 
highest momenta in the $B$ decay final states.

We reconstruct the  primary photon, originating from the $B$ decay 
candidate, using an EMC shower not associated 
with a track. We require that the photon candidate fall within the fiducial 
region of the EMC, has the expected lateral shower shape, and is well-separated
from other  tracks and  showers in the EMC. The primary photon energy, 
calculated in the  \UfourS\ frame, is required to be  in the range $1.6$
-- $2.7$  \gev. We veto photons from $\pi^0$($\eta$) decays  by
requiring that the invariant mass of the primary photon candidates 
combined with any other photon candidate of laboratory energy greater
than $50$ ($250$) 
\mev\ not  be within the range $115$-$155$  ($507$-$587$)  \mevcc. 
Charged $K$ candidates are selected from  tracks, by using  particle 
identification from the DIRC and the \dedx\ measured in the SVT and DCH.

The $B$ decay daughter candidates are reconstructed through their decays
$\piz\ra\gaga$, $\eta\ra\gaga$ (\etagg), $\eta\ra\pip\pim\piz$
(\etappp), $\etapr\ra\etagg\pip\pim$ (\etapepp), and 
$\etapr\ra\rhoz\gamma$ (\etaprg), where $\rhoz\ra\pip\pim$.
Here we require the laboratory energy of the photons to be  greater 
than 50 \mev\ 
(200 \mev\ for  \etaprg). 
We impose the following requirements on the invariant mass in \mevcc\ of 
these particles' final states:
$120 < m(\gamma\gamma) < 150$ for \piz,  $490 < m(\gamma\gamma) < 600$
for \etagg,  
$520 < m(\pip\pim\piz) < 570$ for   \etappp, $930 < m(\pip\pim\eta)
<990$ for \etapepp,  
$910 < m(\pip\pim\gamma) <1000$ for \etaprg, and   $510 < m(\pip\pim)
<1000$ for \rhoz.    
For the \etapr\ and $\eta$ these requirements are 
sufficiently loose as to include sidebands, since these observables are used 
in the
maximum-likelihood (ML) fit described below.  Secondary
pions in \etapr\ and $\eta$ candidates are rejected if their DIRC
and \dedx\ signatures  satisfy tight requirements for being consistent
with protons, kaons, or electrons.  

Neutral $K$ candidates are formed from pairs of oppositely-charged
tracks with a vertex 
$\chi^2$ probability larger than $0.001$,  $486 < m(\pip\pim)<510$
\mevcc\  and a reconstructed decay length greater than
three times  its uncertainty.
We require the momentum of the $\eta$ or \etapr\ in the  \UfourS\
frame to be greater than $0.9$ \gevc\ 
($0.6$ \gevc\ in modes with \etapepp). The invariant mass of $\eta K$
and $\etapr K$ systems 
is required to be less than 3.25 \gevcc.
In $\etapr K \gamma$ final states, we suppress  background from the
decay   $\jpsi K$,  with 
 $\jpsi\ra \etapr \gamma$  by applying a veto on the 
 reconstructed   $\etapr \gamma$   invariant mass. 
Defining the helicity frame for a meson as its rest
frame with polar axis along the direction of the boost from the parent
rest frame, and the decay angle $\theta_{\rm dec}$ as the polar angle of
a daughter momentum in this helicity frame, we require for the \etaprg\
decays $|\cos\theta^{\rho}_{\rm dec}| < 0.9$, and for \etagg\ decays 
 $|\cos{\theta^{\eta}_{\rm dec}}| < 0.9$, to suppress combinatorial background.

A $B$ meson candidate is reconstructed by combining  an $\eta$
or \etapr\ candidate, a charged or neutral kaon and a primary photon
candidate. It is characterized kinematically by the energy-substituted
mass $\mes \equiv \sqrt{(s/2 + \pvec_0\cdot \pvec_B)^2/E_0^2 - \pvec_B^2}$ and
energy difference $\DE \equiv E_B^*-\half\sqrt{s}$, where the subscripts $0$ and
$B$ refer to the initial \UfourS\ and to the $B$ candidate in the lab-frame, respectively,
and the asterisk denotes the \UfourS\ frame.

Background arises primarily from random track combinations in
$\epem\ra\qqbar$ events.
 We reduce this background by using the angle
\thetaT\ between the thrust axis of the $B$ candidate in the \UfourS\
frame and the thrust axis of the rest of the event.
The distribution of $|\costhr|$ is
sharply peaked near $1$ for combinations drawn from jet-like \qqbar\
events, and is nearly uniform for  \BB\ events.  We require
$|\costhr|<0.9$. Furthermore events should contain at least the number
of charged  
tracks in the candidate  decay mode plus one. For \fetaggkpgamma\
we require at least 3 charged tracks in the event.
If an 
event has  multiple $B$ candidates, we select the  candidate  with 
the highest $B$ vertex $\chi^2$ probability.

We obtain signal event yields from unbinned extended maximum-likelihood
fits.  The principal input observables are \DE, \mes\ and a Fisher 
discriminant \xf. Where  relevant, 
the invariant masses \mres\ 
of the intermediate resonances and  
$|\cos\theta^{\rho}_{\rm dec}|$ 
are also used.
The Fisher discriminant \xf\  combines four variables: the angles with
respect to the  
beam axis of the $B$ momentum and the $B$ decay product's thrust axis 
(in the \UfourS\ frame), and the zeroth and second angular moments $L_{0,2}$ 
of the energy flow about the \Bz\ thrust axis.  The moments are defined by
$ L_j = \sum_i p_i\times\left|\cos\theta_i\right|^j$,
where $\theta_i$ is the angle with respect to the $B$ thrust axis of
track or neutral cluster $i$, $p_i$ is its momentum, and the sum
excludes the $B$ candidate daughters.

We estimate \BB\ backgrounds using simulated samples of $B$
decays~\cite{geant}. Signal   
and inclusive $b\ra s\gamma$ events are simulated according to the
Kagan-Neubert model~\cite{KN}.  
The branching fractions in the simulation are based on measured 
values and theoretical predictions.

For each event $i$ and hypothesis $j$ (signal,  continuum  or \BB\ background),
the likelihood function is
\begin{equation}
{\cal L}=  e^{-\left(\sum n_j\right)} \prod_{i=1}^N \left[\sum_{j=1}^3 n_j  {\cal P}_j ({\bf x}_i)\right] ,
\end{equation}
where  $N$ is the number of input events,  $n_j$ is the number of events for 
hypothesis $j$ and     ${\cal P}_j ({\bf x}_i)$ is the corresponding
probability density function (PDF), evaluated   
with the observables ${\bf x}_i$ of the $i$th event. 
Since   correlations among the observables are small,
we take each  ${\cal P}$  as the product of the PDFs for the separate
variables. 
We determine the PDF parameters from Monte Carlo simulation for the
signal and \BB\ background, while using 
   sideband data ($5.25 < \mes\
<5.27$ \gevcc; $0.1<|\DE |<0.2$ \gev )  
to model the PDFs of  continuum  background. We parameterize each of 
the functions ${\cal P}_{\rm sig}(\mes),\ 
{\cal  P}_{\rm sig}(\DE),\ { \cal P}_j(\xf),\ $ and the 
components of ${\cal P}_j(\mres)$ that peak in \mes\ with
either a Gaussian, the sum of
two Gaussian distributions, or an asymmetric Gaussian function, as
required, to describe the  
distribution.  Distributions of  \DE\ for \BB\ and continuum background and  
$|\cos\theta^{\rho}_{\rm dec}|$ are represented by linear or 
quadratic functions.
The \BB\ and continuum background in \mes\ is described by the ARGUS function 
$x\sqrt{1-x^2}\exp{\left[-\xi(1-x^2)\right]}$, with
$x\equiv2\mes/\sqrt{s}$ and a  
parameter $\xi$~\cite{argus}. We allow  continuum background PDF
parameters  to vary in the fit.

Large control samples of $B$ decays to charmed final states of similar 
topology are used to verify the simulated resolutions in \DE\ and \mes.
Where the control data samples reveal differences from the Monte Carlo (MC) in mass or energy
resolution, we shift or scale the resolution used in the likelihood fits.
Any bias in the fit is determined from a large set of simulated
experiments in which the \qqbar\ background is generated from the PDFs, and
into which we have embedded  
the expected number of \BB\  background  and signal events 
chosen  randomly from fully 
simulated Monte Carlo samples. 

\begin{table*}[htp]
\caption{
Signal yield, detection
efficiency  \eff, daughter branching fraction product $\prod\calB_i$,
significance \signf 
(including systematic uncertainties), measured branching
fraction \calB\ with statistical error for each decay mode. For the
combined measurements we give
the significance  (with systematic uncertainties included) and the  branching fraction
with statistical and systematic uncertainty (in parentheses the  90\%
C.L. upper limit). For the \fetakpgamma\ mode we also list the measured  signal charge asymmetry \acp.
}
\label{tab:results}
\begin{tabular}{lcccccc}
\dbline
Mode& \quad Yield \quad&\quad \eff \quad &\quad
$\prod\calB_i$ (\%) \quad&\quad \signf \quad &\quad \bfemsix \quad & \acp\ ($10^{-2}$) \\
\tbline
~~\fetaggkzgamma &   $36^{+13}_{-12}$ &$10.2$&$13.6$&$4.6$&$11.2^{+4.0}_{-3.7}$&\\
~~\fetatrepikzgamma &   $14^{+8}_{-7}$ &$7.0$&$7.8$&$2.9$&$11.5^{+6.1}_{-5.3}$&\\
\bma{\fetakzgamma}&  &  &  &{\bma \setaKzg}& {\bma \RetaKzg} &   \\
\tbline
~~\fetaggkpgamma &   $110^{+22}_{-21}$  &$12.9$&$39.4$&$8.0$&$9.4^{+1.8}_{-1.7}$&$-1.3 \pm 15.3$ \\ 
~~\fetatrepikpgamma &   $53^{+14}_{-13}$  &$8.8$&$22.6$&$6.6$&$11.4^{+3.0}_{-2.8}$&$-21.9 \pm 20.5$ \\
\bma{\fetakpgamma} &  &  &  &{\bma \setaKpg} & {\bma \RetaKpg } &${\bf -8.6 \pm 12.0\pm 1.0}$\\
\tbline
~~\fetaprEppkzgamma &$1^{+2}_{-2}$&$6.2$& $6.0$&$0.4$&$0.6^{+2.8}_{-2.0}$&   \\
~~\fetaprRgkzgamma &$14^{+16}_{-14}$&$5.3$& $10.2$&$1.2$&$11.2^{+12.8}_{-11.0}$ &  \\
\bma{\fetaprkzgamma} &  &  &  &{\bma \setapKzg} &{\bma \RetapKzg} ({\bma < \uletapKzg}) & \\
\tbline
~~\fetaprEppkpgamma &  $6^{+6}_{-5}$  &$8.2$&$17.5$&$1.6$&$1.9^{+1.8}_{-1.4}$&$$ \\
~~\fetaprRgkpgamma &  $10^{+27}_{-24}$  &$9.9$&$29.5$&$0.5$&$1.5^{+3.9}_{-3.6}$ &$$\\
\bma{\fetaprkpgamma}  &  &  &  &{\bma \setapKpg} &{\bma \RetapKpg} ({\bma < \uletapKpg}) &  \\
\dbline
\end{tabular}
\vspace{-5mm}
\end{table*}

In Table \ref{tab:results} we show the signal yield, the efficiency, 
and the product of daughter branching fractions for each decay mode. 
The efficiency is calculated as the ratio of the number of signal MC
events entering 
 into the ML fit to the total generated.  
We compute the branching fractions from the fitted  signal event
yields, reconstruction efficiencies, 
daughter branching fractions, and the number of produced $B$ mesons,
assuming equal production  
rates of charged and neutral $B$ pairs. We correct the yield for any
bias estimated by the simulations. 
We combine results from different channels by combining their
likelihood functions,  taking into  account
 the correlated and uncorrelated systematic errors.
We report the statistical significance and 
branching fraction for the individual decay channel; for 
combined measurements having a significance smaller than 5$\sigma$, we 
also report
the 90\% confidence level  (C.L.) upper limit. 

The statistical error on the signal yield is taken as the change in 
the central value when the quantity $-2\ln{\cal L}$ increases by one 
unit from its minimum value. The significance is  the square root 
of the difference between the value of $-2\ln{\cal L}$ (with systematic 
uncertainties included) for zero signal and the value at its minimum.
The 90\%  C.L. upper limit is taken to be the branching 
fraction below which lies 90\% of the total  likelihood integral 
in the positive branching fraction region.

The measured  charge asymmetry in the decay  \etaKpg\ is 
corrected for an estimated bias of  $-0.005\pm0.010$, determined from 
studies of signal Monte Carlo events and data control samples and from 
calculation of the asymmetry due to particles interacting in the 
detector. The result is $\acp = \aetaKpg\ $  
with an asymmetry interval [$-0.282,0.113$]  at 90\% C.L..

\begin{figure}[tb]
 \includegraphics[angle=0,scale=0.4335]{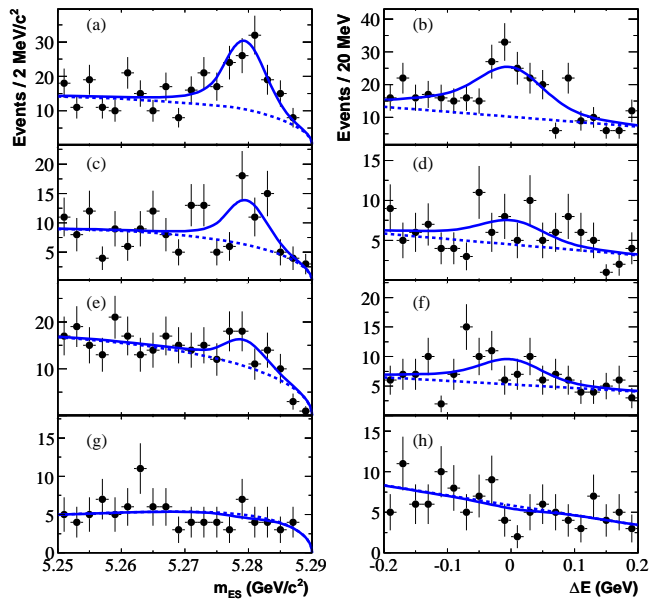}
\vspace{-0.7cm}
 \caption{\label{fig:projections}
 The $B$ candidate \mes\ and \DE\ projections for \fetakpgamma\ (a, b),
 \fetakzgamma\ (c, d), \fetaprkpgamma\ (e, f) and \fetaprkzgamma\ (g, h).
 Points with error bars represent the data, the solid
line the full fit function, and the dashed line its background component.}
\vspace{-.2cm}
\end{figure}

Figure~\ref{fig:projections} shows, as representative fits, 
the  projections onto \mes\ and \DE\ 
for the decays \fetakpgamma ,   \fetakzgamma ,
 \fetaprkpgamma\    and \fetaprkzgamma\ for
a subset of the data for which the signal likelihood
(computed without using the variable plotted) exceeds a mode-dependent
threshold that optimizes the sensitivity.

\begin{figure}[t]
\resizebox{\columnwidth}{!}{
\begin{tabular}{cc}
\includegraphics[]{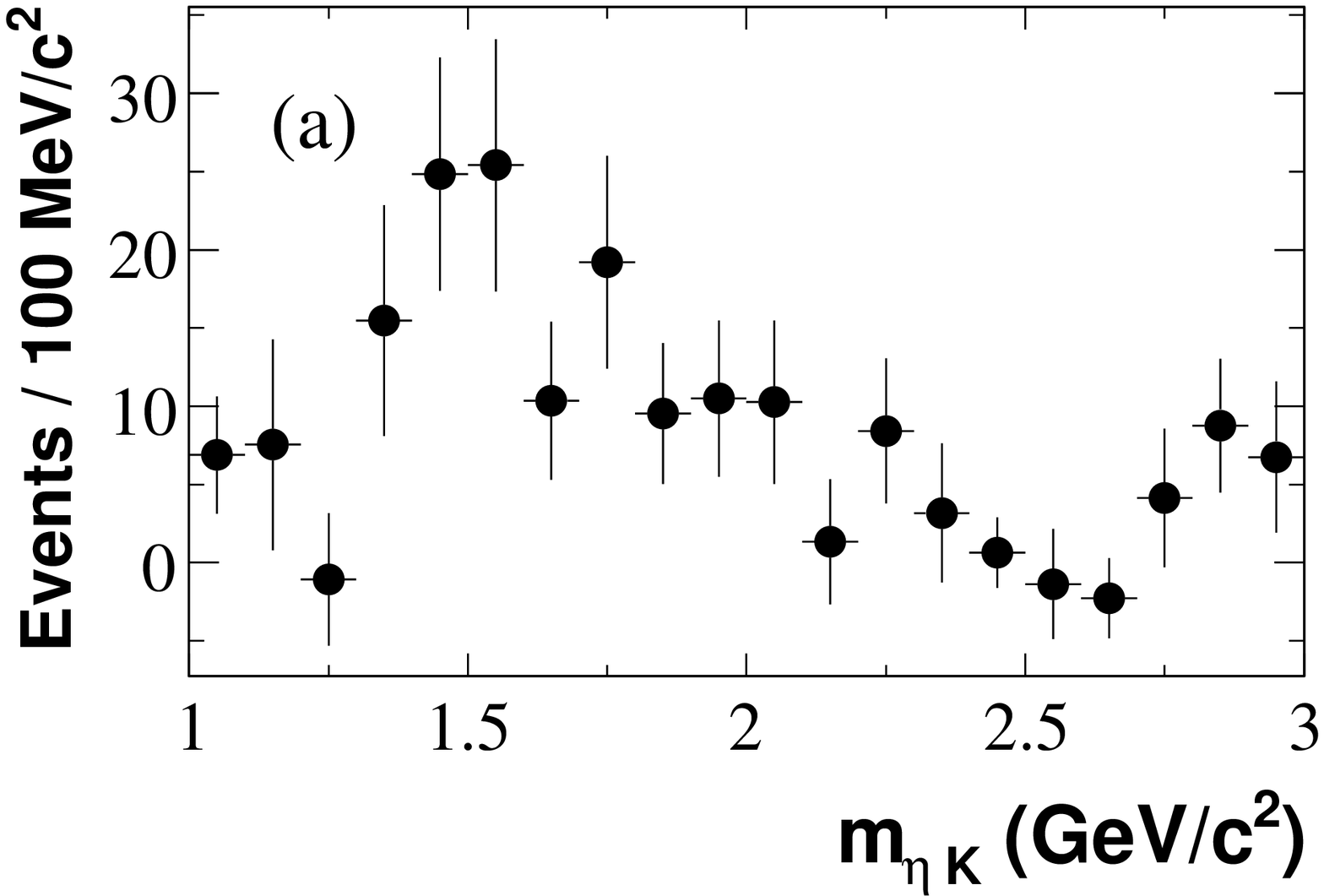} &
\includegraphics[]{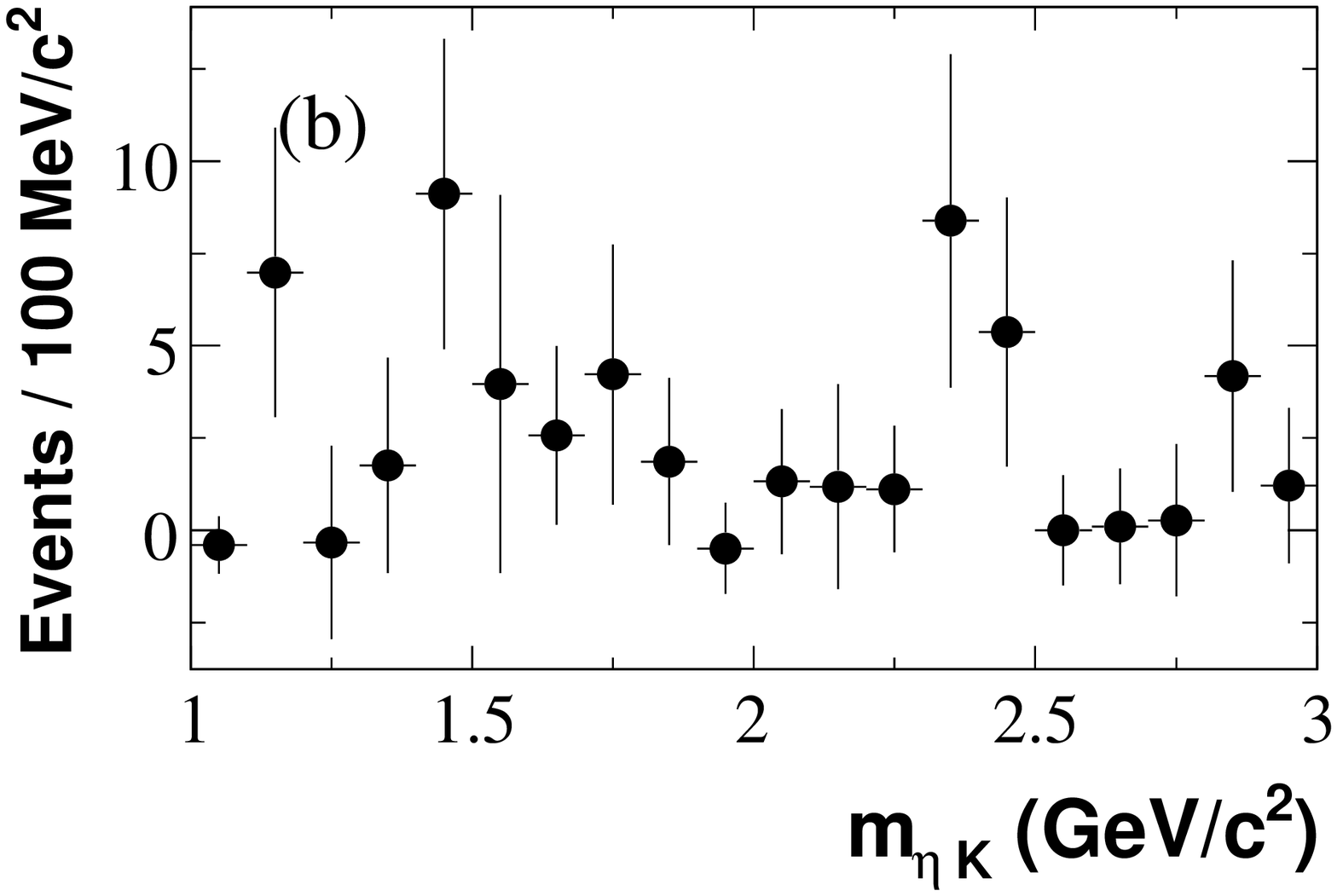} 
\end{tabular}
}
\caption{Plot of $\eta K$ invariant mass for signal using the
  weighting technique described in the text for the combined sub-decay
  modes: (a) charged modes,  (b) neutral modes.}
\label{fig:sPlots}
\end{figure}

Figure~\ref{fig:sPlots} shows the distribution of the $\eta K$
invariant mass for signal events obtained by 
the event-weighting technique (sPlot) described in Ref.~\cite{sPlots}.
We use the covariance matrix and PDFs
from the ML fit  to determine a probability for each signal event.
  The resulting distributions (points with errors) 
are normalized to the signal yield. This mass distribution is useful
to compare with theoretical predictions for radiative decays~\cite{KN}.

The main sources of systematic error include uncertainties in the
 PDF  parameterization  and ML fit bias.    For the
signal, the uncertainties in PDF parameters are estimated by comparing
MC and data in control samples.  Varying the 
signal PDF parameters within these errors, we estimate yield
uncertainties of 1--2 events, depending on the mode.  The uncertainty
from fit bias is taken as half the correction
itself (1--3 events). 
Systematic uncertainties due to lack of knowledge of the primary photon
spectrum are estimated to be in the range 2-6\% depending on the decay mode. 
Uncertainties in our knowledge of the efficiency, found from auxiliary
studies, include $0.8\%\times N_t$ and $1.5\%\times N_\gamma$, where
$N_t$ and $N_\gamma$ are the numbers of tracks and photons, respectively,
in the $B$ candidate. There is a  systematic error of 2.1\% in the
efficiency of \KS\ reconstruction. 
  The uncertainty in the total number of \BB\ pairs in the
data sample is 1.1\%.  Published data~\cite{PDG2004}\ provide the
uncertainties in the $B$ daughter product branching fractions (0.7-3.4\%).
We assign a systematic uncertainty of 0.010 to \acp\ for the bias correction.

In conclusion, we have measured the central values  and 90\% C.L.  upper 
limits in units of $10^{-6}$ for the branching fractions: 
$\BretaKzg = \RetaKzg\  $,
$\BretaKpg = \RetaKpg\   $,
$\BretapKzg = \RetapKzg\   (<\uletapKzg)$,
$\BretapKpg = \RetapKpg\   (<\uletapKpg)$.
The measured branching fractions of the decay modes \etaKpg\ and \etaKzg\ are
in good agreement with the values reported by the Belle
Collaboration~\cite{KEta}. 
We do not find evidence  of the decays  \etapKzg\ and  \etapKpg.
The   \BetapKGamma\ decays may be suppressed 
with respect to   \BetaKGamma\  decays  due to destructive
interference between 
two penguin amplitudes~\cite{Lipkin}. This effect has been measured
in $B$ decays to $\etapr K$  and $\eta K$, for which the branching
fraction of the 
former is enhanced with respect to that of the latter~\cite{EtaPK}.
We have also measured the  charge asymmetry in the decay \etaKpg\ to be \acp =
\aetaKpg, 
consistent with zero. The \acp\  interval at 90\% C.L. is  [$-0.28,0.11$].

We are grateful for the excellent luminosity and machine conditions
provided by our \pep2\ colleagues, 
and for the substantial dedicated effort from
the computing organizations that support \babar.
The collaborating institutions wish to thank 
SLAC for its support and kind hospitality. 
This work is supported by
DOE
and NSF (USA),
NSERC (Canada),
IHEP (China),
CEA and
CNRS-IN2P3
(France),
BMBF and DFG
(Germany),
INFN (Italy),
FOM (The Netherlands),
NFR (Norway),
MIST (Russia), and
PPARC (United Kingdom). 
Individuals have received support from CONACyT (Mexico), 
Marie Curie EIF (European Union),
the A.~P.~Sloan Foundation, 
the Research Corporation,
and the Alexander von Humboldt Foundation.


\begin{thebibliography}{99}
\bibitem{ChargeCon}
Charge-conjugate modes are implied throughout.
\end{thebibliography}
\end{document}